# Performance of TCP/IP Using ATM ABR and UBR Services over Satellite Networks[1]

Shiv Kalyanaraman, Raj Jain, Rohit Goyal, Sonia Fahmy
Department of Computer and Information Science
The Ohio State University
Columbus, OH 43210-1277
E-mail: {*shivkuma, jain, goyal, fahmy* }@cis.ohio-state.edu
and
Seong-Cheol Kim
Principal Engineer, Network Research Group
Communication Systems R&D Center
Samsung Electronics Co. Ltd.
Chung-Ang Newspaper Bldg.
8-2, Karak-Dong, Songpa-Ku
Seoul, Korea 138-160
Email: kimsc@metro.telecom.samsung.co.kr

**Abstract**

We study the buffering requirements for zero cell loss for TCP/IP over satellite links using the available bit rate (ABR) and unspecified bit rate (UBR) services of asynchronous transfer mode (ATM) networks. For the ABR service, we explore the effect of feedback delay (a factor which depends upon the position of the bottleneck), the switch scheme used, and background variable bit rate (VBR) traffic. It is shown that the buffer requirement for TCP over ABR is independent of the number of TCP sources, but depends on the aforementioned factors. For the UBR service, we show that the buffer requirement is the sum of the TCP receiver window sizes. We substantiate our arguments with simulation results.

# 1 Introduction

ATM networks offer two classes of service for data traffic: available bit rate (ABR) and unspecified bit rate (UBR). In ABR, the sources constantly monitor the network feedback and adjust their load on the network. In UBR, there is no explicit feedback. The network simply drops packets (or cells) if it is overloaded. Thus, UBR users may use their own load control algorithm to minimize the loss under network overload. Since TCP does have such a load control mechanism, many believe that the extra complexity of ABR is not required. There is currently a debate between supporters of UBR and ABR.

In general the efficiency of networking protocols is a decreasing function of the propagation delay. Satellite links have an extremely long delay and, therefore, many protocols do not

---

perform well on paths including satellite links. It is important to verify how the two ATM data services perform on satellite links.

TCP is the most popular transport protocol for data transfer. It provides a reliable transfer of data using a window-based flow and error control algorithm [7]. When TCP runs over ABR, there are two control algorithms active: the TCP window-based control running on top of the ABR rate-based control. It is, hence, important to verify that the ABR control performs satisfactorily for TCP/IP traffic.

Since the ABR control was recently standardized, it has not been extensively studied, especially in the context of satellite links. The behavior of TCP over the Unspecified Bit Rate (UBR) service class using LAN and WAN links have been studied in [8, 9, 10, 11, 2]. The UBR class [6] is the lowest priority class in ATM. UBR does not include flow control and depends upon transport layers to provide flow control. The only degree of freedom to control traffic in UBR is through buffer allocation (which includes cell drop policies).

When compared to UBR, ABR has additional degrees of freedom in the switch schemes and source parameters. ABR performs better than UBR even when the ATM switches use simple binary (EFCI) feedback schemes [8, 9]. When switches give explicit rate feedback, given adequate buffering, TCP can achieve maximum throughput with zero loss [5]. The buffering requirements for zero loss TCP over ABR and UBR have been studied in the LAN and WAN environments [4, 3, 2].

In this paper, we study the buffer requirements for zero loss TCP over ABR and UBR networks having satellite links. The ABR networks use Explicit Rate (ER) switches running the ERICA algorithm [1]. The application traffic on top of TCP is assumed to be unidirectional and infinite (eg, a large file transfer). Traffic in satellite networks may be bottlenecked before reaching the satellite link or may be bottlenecked after traversing the satellite link. We characterize the bottleneck position by the sum of the time taken for feedback to reach from the bottleneck to the source and the time taken for the effect of the new load to reach the bottleneck switch. This quantity is called the *feedback delay*. In this paper, we study the effect of feedback delay on the amount of buffers required. We also study the effect of the switch scheme used and the effect of varying ABR capacity on the buffer requirement.

For TCP running over UBR, the position of the bottleneck is not important because there is no explicit feedback possible before one round trip. We show that the buffers required is equal to the sum of the receiver window sizes of the TCP connections.

## 1.1 The ERICA Switch Scheme

In this section, we present a brief overview of the ERICA switch algorithm. More details can be found in [1].

Explicit Rate Indication for Congestion Avoidance (ERICA) is a simple switch scheme that allocates bandwidth fairly with a fast response. The scheme consists of using a Target Utilization of, say, 90%. The Target Rate is then set at:

$$Target\ Rate = Target\ Utilization \times Link\ Rate$$

Since VBR and CBR are serviced first, bandwidth available for ABR service class is given by:

$$ABR = Target\ Rate\ \text{-}\ VBR\ \text{-}\ CBR$$

The overload is measured with respect to the target rate (and not link rate):

$$Overload = Input\ Rate\ /\ ABR$$

In addition to the input rate, the switches also measure the number of active virtual connections (VCs) and compute the fair share:

$$Fair\ Share = ABR\ /\ Number\ of\ Active\ VCs$$

For each VC, its share is computed based on the overload factor and the VC's current cell rate:

$$A\ VC's\ Share = VC's\ Current\ Cell\ Rate\ /\ Overload$$

The VC is given the maximum of its share as computed above or the fair share.

$$ER\ for\ VC = max\ (Fair\ Share,\ VC's\ Share)$$

The explicit rate (ER) in the RM cell is reduced if ER for VC as computed above is less:

$$ER\ in\ Cell = min\ (ER\ in\ Cell,\ ER\ for\ the\ VC)$$

This simple algorithm has several desirable properties including fast response time, low queue length, and simplicity. We have enhanced this algorithm to reduce spiking effects due to transient overloads. We use the enhanced version in our simulation. The essential aspects of the study however remain the same for both versions of ERICA.

To show the effect of different switch schemes, we also use an improved version of ERICA, called ERICA+. Whereas ERICA uses only the input load as the congestion metric, ERICA+ uses a combination of input load and the queueing delay as the congestion metric to allow better control of queues.

## 2  TCP And ERICA Options

We experiment with an infinite TCP source running on TCP over an ATM WAN[2]. The TCP source always has a frame to send. However, due to TCP window constraint, the resulting traffic at the ATM layer may or may not be continuous. We use a TCP maximum segment size (MSS) of 512 bytes. The window scaling option is used so that the throughput is not limited by the path length. The TCP window is set at $34000 \times 2^8$ (using the window scaling option). This window size is sufficient for a single TCP source to fill a 550 ms (one round trip) pipe.

---

[2]Note that the effect of bursty TCP sources is yet to be studied. All results presented here assume a TCP source which never runs out of packets to send. It is expected that with bursty sources, queue lengths will be larger than those indicated here.

The zero-loss buffer requirement applies for all TCP congestion algorithms including "fast retransmit and recovery" algorithms. These algorithms are equivalent since there is no packet loss (assuming that spurious timeouts do not occur).

The ERICA algorithm uses two key parameters: target utilization and averaging interval length. The algorithm measures the load and number of active sources over successive averaging intervals and tries to achieve a link utilization equal to the target. The averaging intervals end either after the specified length or after a specified number of cells have been received, whichever happens first. In the simulations reported here, the target utilization is set at 90%, and the averaging interval length defaults to 100 ABR input cells or 1 ms, represented as the tuple (1 ms, 100 cells).

## 3 The $n$ Source + VBR Configuration

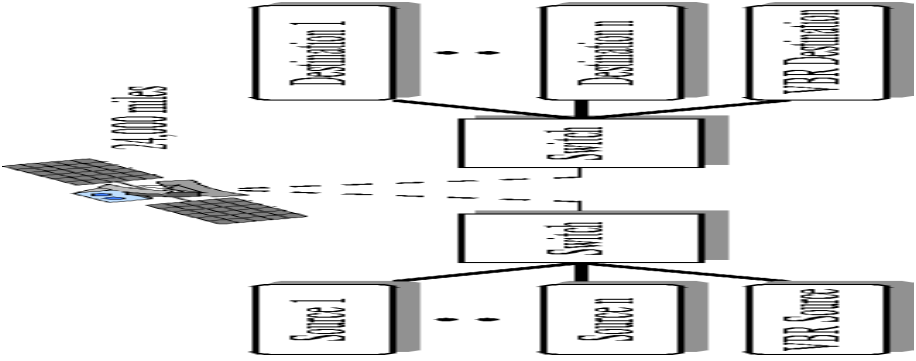

Figure 1: $n$ Source + VBR Configuration

The general configuration we analyze is what we call "the $n$ Source + VBR configuration." This configuration has a single bottleneck link (the satellite link) shared by the $n$ ABR sources and possibly a VBR source. All links run at 155 Mbps. The total round trip time is 550 ms. The feedback delay, twice the propagation delay from the sources to the first switch may assume values 0.01 ms, 1 ms, 10 ms and 550 ms.

The VBR background is optional. When present, it is an ON-OFF source with a 1 ms ON time and 1 ms OFF time. The VBR starts at t = 2 ms. The maximum amplitude of the VBR source is 124.41 Mbps (80% of link rate). VBR is given priority at the link, i.e, if there is a VBR cell, it will be scheduled for output on the link before any waiting ABR cells are scheduled.

All traffic is unidirectional. A large (infinite) file transfer application runs on top of TCP for the TCP sources. The variable $n$ may assume values 5 and 15.

Our performance metric is the maximum queue length. Observe that the Round Trip Time (RTT) of the network measured in cells is 550 ms × 365 cells/ms = 200,750 cells. We characterize the buffering as a fraction of RTT.

# 4 Simulation Results

We first present the results with short feedback delays. These results show that the buffer requirement depend upon the feedback delay. Next we make the feedback delay large and show that the buffer requirement is quite large. Under these circumstances, the switch scheme used can significantly reduce the buffer requirement. The third set of experiments shows the effect of VBR traffic which causes high variance in the ABR capacity. Again the improved switch scheme (ERICA+) can control the queues to desired values.

The last subsection shows that the queues with TCP over UBR depend upon the sum of the TCP receiver windows.

## 4.1 Effect of Number of Sources and Feedback Delay

Table 1 shows the effect of the number of sources and feedback delay on the maximum queue length. Notice that the queue length increases when either the number of sources or the feedback delay is increased. However, the increase is not proportional in either case.

## 4.2 Effect of Long Feedback Delays and the Switch Scheme

Table 2 shows the maximum queues when the feedback delay is 550 ms (equal to the RTT). In this case, the basic ERICA algorithm does not does not converge within the simulation time (about 20 round trips). The queue length keeps increasing (denoted by "unbounded" in the table). This is because TCP itself takes 15 round trips to load the link and the remaining time is not enough for the switch scheme to converge given the long RTT and the varying demand of the TCP sources. The queue, however, stabilizes for the enhanced scheme, ERICA+ at 1.6 × RTT. The large buffer requirement is due to the long feedback delay of this configuration. Thus, we conclude that the switch scheme has a significant effect on the buffer requirements for the ABR service.

Table 1: Effect of Number of Sources and Feedback Delay

| Number of Sources | Feedback Delay (ms) | Max Q (cells) |
|---:|---:|---:|
| 5 | 0.01 | 1229 (0.006 × RTT) |
| 15 | 0.01 | 2059 (0.01 × RTT) |
| 5 | 10 | 18356 (0.09 × RTT) |
| 15 | 10 | 17309 (0.086 × RTT) |

Table 2: Effect of Long Feedback Delays and the Switch Scheme

| Number of Sources | Feedback Delay (ms) | Switch Scheme | Max Q (cells) |
|---:|---:|:---:|:---:|
| 15 | 550 | ERICA | Unbounded |
| 15 | 550 | ERICA+ | 1.6 × RTT |

## 4.3 Effect of High Frequency VBR Background Traffic

Table 3 shows the maximum queues when a high frequency VBR background is used. We also vary the feedback delay parameter (0.01 ms and 10 ms) and the switch scheme used (ERICA and ERICA+). We find that the basic ERICA scheme does not control the queues due to the high variance in the workload. However, ERICA+ controls the queues to small values in both its experiments.

## 4.4 TCP over UBR Satellite Links

We find in our experiments that, irrespective of the position of the bottleneck, the UBR buffer requirement when $n = 5$ is 817819 cells. The sum of the (five) TCP windows is 821133 cells. This extends our earlier observation [2] that the buffer requirements for UBR equal the sum of the TCP receiver windows.

Table 3: Effect of High Frequency VBR Background Traffic

| Number of Sources | Feedback Delay (ms) | Switch Scheme | Max Q (cells) |
|---|---|---|---|
| 15+VBR | 0.01 | ERICA | Unbounded |
| 15+VBR | 10 | ERICA | Unbounded |
| 15+VBR | 0.01 | ERICA+ | 2006 (0.01 × RTT) |
| 15+VBR | 10 | ERICA+ | 5824 (0.028 × RTT) |

# 5 Summary

We have study the buffering requirements for zero cell loss for TCP over Satellite ATM networks using the ABR and UBR service. The buffer requirement for TCP over satellite ABR networks is independent of the number of TCP sources, but depends on factors such as the feedback delay (twice the propagation delay from the bottleneck to the source), the switch scheme used and the background VBR traffic. For the UBR service, we show that the buffer requirement is the sum of the TCP receiver window sizes.

---

[3] Throughout this section, AF-TM refers to ATM Forum Traffic Management sub-working group contributions.

[4] All our papers and ATM Forum contributions are available through http://www.cis.ohio-state.edu/~jain